\documentclass[aps,prd,preprint,showpacs,superscriptaddress]{revtex4-1}
\usepackage{amsmath}
\usepackage{latexsym}
\usepackage{amsfonts}
\usepackage{graphicx}
\usepackage{mathrsfs}
\usepackage{epstopdf}
\usepackage{ulem}

\begin{document}

\title{Dynamical analysis of modified gravity with nonminimal gravitational
coupling to matter}

\author{Rui An}
\email{an\_rui@sjtu.edu.cn}
\affiliation{Department of Physics and Astronomy, Shanghai Jiao Tong University,
Shanghai 200240, China}

\author{Xiaodong Xu}
\email{xiaodong.xu@uct.ac.za}
\affiliation{Astrophysics, Cosmology and Gravity Centre, Department of Mathematics and Applied Mathematics, University of Cape Town, Rondebosch 7701, Cape Town, South Africa}

\author{Bin Wang}
\email{wang\_b@sjtu.edu.cn}
\affiliation{Department of Physics and Astronomy, Shanghai Jiao Tong University,
Shanghai 200240, China}

\author{Yungui Gong}
\email{yggong@mail.hust.edu.cn}
\affiliation{School of Physics, Huazhong University of Science and Technology,
Wuhan, Hubei 430074, China}

\begin{abstract}
We perform a phase space analysis of a generalized modified gravity
theory with nonminimally coupling between geometry and matter. We
apply the dynamical system approach to this generalized model and
find that in the cosmological context, different choices of Lagrangian
density will apparently result in different phases of the Universe.
By carefully choosing the variables, we prove that there is an attractor
solution to describe the late time accelerating universe when the
modified gravity is chosen in a simple power-law form of the curvature
scalar. We further examine the temperature evolution based on the
thermodynamic understanding of the model. Confronting the model with
 supernova type Ia data sets, we find that
the nonminimally coupled theory of gravity is a viable model to describe
the late time Universe acceleration.
\end{abstract}

\pacs{98.80.Cq, 98.80.-k, 04.50.Kd}
\preprint{1512.xxxxx}

\maketitle

\section{Introduction}

Recent astronomical observations consistently tell us that our universe
is undergoing accelerating expansion. One possible explanation of this
accelerating expansion is the standard $\Lambda$ cold dark matter ($\Lambda$CDM)
cosmological model in the framework of Einstein gravity, containing
dominant dark energy responsible for the cosmic acceleration. Another
promising way to explain the accelerating expansion of the universe is
to assume that at large scales Einstein gravity breaks down, and
more general actions are needed to describe the gravitational
field. One of these generalized actions is to replace the standard Einstein-Hilbert
action with an arbitrary nonlinear function of the Ricci scalar, which
is called $f(R)$ gravity. $f(R)$ generalized gravity models have
been extensively investigated lately, for a review please see for
example \cite{1Sotiriou2010 ,2Felice2010,3Nojiri2011} and references
therein.

The action of the $f(R)$ modified gravity theory can be further generalized
by introducing the nonminimal coupling between
matter and geometry in the action \cite{4Bertolami2007}. The action
of this gravity theory is of the form
\begin{equation}
\label{action1}
S=\int d^{4}x\sqrt{-g}\left\{\kappa f_{1}(R)+[1+\lambda f_{2}(R)]\mathcal{L}\right\},
\end{equation}
where $\kappa=c^{4}/(16\pi G)$, $R$ is the Ricci scalar curvature,
$f_{1}(R)$ and $f_{2}(R)$ are arbitrary functions of $R$, $\mathcal{L}$
is the matter Langrangian density, and $\lambda$ is the coupling
constant determining the strength of the interaction between $f_{2}(R)$
and the matter Lagrangian. Taking $f_{1}(R)=R-2\varLambda$ and $\lambda=0$,
\eqref{action1} returns to the standard Einstein-Hilbert action. This nonminimal
coupling model leads to considerable cosmological implications, such
as the solar system \cite{5Bertolami2013} and stellar dynamics\cite{6Bertolami3Paramos1},
mimicking of dark matter by leading to the flattening of the galaxy
rotation curves \cite{7Bertolami2009}, the modeling
of the cosmic speed up at late times \cite{8Bertolami2010,9Bisabr2012},
the reheating scenario after inflation \cite{10Fakir1990} and the
virial equilibrium in galaxy cluster dynamics \cite{11Futamase1989},
etc.

In Einstein gravity, different forms of  matter
Lagrangian densities like $\mathcal{L}=p$ and $\mathcal{L}=-\rho$ are
perfectly equivalent, which lead to the same
consequences in the gravitational field equations and the conservation
of the energy-momentum. However in the generalized gravity model where
matter is coupled nonminimally with the scalar curvature, different
Lagrangian densities which are classically equivalent do not yield
the same gravitational field equations \cite{12Bertolami2008}.
Depending on the nature of the matter Lagrangian this may lead to the appearance of an extra force on the matter particles which makes the motion of particles nongeodesic, but with an unchanged continuity equation when $\mathcal{L}=-\rho$. Or when the matter Lagrangian has the form $\mathcal{L}=p$, there is no extra force on the particles, which implies that the matter particles follow the geodesics of the background metric, while the standard continuity equation no longer holds and the matter energy is not conserved \cite{13Bertolami2013}.

Recently it was argued that a nonminimal coupling between the scalar
curvature and the matter Lagrangian density may account for the accelerating
expansion of the Universe \cite{8Bertolami2010,9Bisabr2012}. In both
papers, the authors adopted the same strategy by assuming
a power-law expansion ansatz $a\sim t^{\beta}$ for the evolution
of the scale factor. They chose for simplicity the gravity in the
forms of $f_{1}(R)=R$ and $f_{2}(R)\sim R^{n}$ and searched for the relation between the exponent $n$ and $\beta$ to
accommodate the cosmic acceleration. Asymptotic accelerated
universe was obtained in their solutions, however the thorough exploration
is difficult since their solutions were based on {\it a priori} of the scale
factor they assumed. The complete match of the accelerated expansion
of the Universe described by the nonminimal coupling with observations
is still lacking. Comparing to finding the solution for the cosmological
acceleration from an assumed priori, the dynamical system approach to
nonminimally coupled $f(R)$ theories is a better way to determine the
solution of the late time acceleration era in general models with
nonminimal coupling between matter and geometry. Attempts in this
direction can be found in \cite{14Ribeiro2014,15Azizi2014}, where
they fixed the matter Lagrangian density
$\mathcal{L}=-\rho$.
It was argued in \cite{14Ribeiro2014} that the late time acceleration
can only be accommodated in the nonminimally coupled $f(R)$ theory
with more complicated assumptions about $f_{1}(R)$ and $f_{2}(R)$. In
\cite{15Azizi2014}, it was confirmed that, assuming $f_{1}(R)=R$
and $f_{2}(R)$ with a general form, or $f_{1}(R)$ to be an arbitrary
function of $R$ and $f_{2}(R)\sim R$, there exists an attractor
solution for the model with nonminimal coupling between
gravity and matter which can explain the late time
accelerating epoch of our Universe. Apart from the success of finding
the late time solutions using dynamical system approach, in most
researches, the effect of the coupling strength between
the matter Lagrangian and gravity on the acceleration is not analyzed.

In this work we will further investigate the cosmological dynamics
of the modified gravity with a nonminimal curvature matter coupling.
We will provide independent analysis in favor of different
Lagrangian densities. In our study, we will fix
$f_{1}(R)=R$ and choose the form of $f_{2}(R)$ to be the power function of $R$.
By appropriately choosing the variables, we will determine the fixed
points of the dynamical system and exhibit the existence of the late
time attractor solution with the simple choice of the modified gravity
to explain the late time acceleration epoch of the Universe. Furthermore
we will examine the radiation temperature evolution in the presence of the geometry-matter
coupling. Employing  the luminosity distance
data from the Union2.1 compilation of type Ia supernovae observations \cite{17Suzuki2012}, we constrain our model parameters. Comparing with the observational data, we will
show the viability of the nonminimally coupled theories of gravity
in the cosmological context.

The organization of the present work is as follows: In Sec. II,
we will briefly introduce the nonminimally coupled modified gravity
model and its field equations. In Sec. III, by
taking particular functional choices of $f_{1}(R)$ and $f_{2}(R)$, we will perform a phase space analysis
on the modified gravity theory with nonminimal curvature-matter coupling. By properly choosing dynamical variables,
we will apply the dynamical
system approach to this generalized gravity model and look for the
attractor solution to account for the late time accelerating
expansion. In Sec. IV, we will examine the temperature
evolution in this nonminimally coupled theory. In the following section,
we will confront the model with  the type Ia supernova data sets. The last section will
be devoted to the conclusion.

\section{The Model}

Variation of the action functional \eqref{action1} for the nonminimally coupled
theory of gravity with respect to the metric yields the gravitational
field equations \cite{13Bertolami2013}
\begin{equation}
\label{modgreq1}
[\kappa F_{1}(R)+\lambda F_{2}(R)\mathcal{L}]R_{\mu\nu}-\frac{\kappa}{2}g_{\mu\nu}f_{1}(R)=\Delta_{\mu\nu}(\kappa F_{1}(R)+\lambda F_{2}(R)\mathcal{L})+[1+\lambda f_{2}(R)]T_{\mu\nu},
\end{equation}
where $F_{i}(R)\equiv f_{i}^{'}(R)$ and the prime denotes the derivative
with respect to the curvature scalar, $\Delta_{\mu\nu}\equiv\nabla_{\mu}\nabla_{\nu}-g_{\mu\nu}\square$,
$T_{\mu\nu}$ is the matter energy-momentum tensor defined as
\begin{equation}
T_{\mu\nu}=-\frac{2}{\sqrt{-g}}\frac{\delta(\sqrt{-g}\mathcal{L})}{\delta g^{\mu\nu}}.
\end{equation}
The trace of \eqref{modgreq1} reads
\begin{equation}
[\kappa F_{1}(R)+\lambda F_{2}(R)\mathcal{L}]R-2\kappa f_{1}(R)=-3\square[\kappa F_{1}(R)+\lambda F_{2}(R)\mathcal{L}]+\frac{1}{2}[1+\lambda f_{2}(R)]T,
\end{equation}
where $T$ is the trace of $T_{\mu\nu}$.
According to the generalized Bianchi identities, we obtain the noncovariant
conservation law,
\begin{equation}
\label{modconseq1}
\nabla^{\mu}T_{\mu\nu}=\frac{\lambda F_{2}(R)}{1+\lambda f_{2}(R)}[g_{\mu\nu}\mathcal{L}-T_{\mu\nu}]\nabla^{\mu}R,
\end{equation}
which can be nonvanishing. This nonconservation can be interpreted as
an energy exchange between geometry and matter \cite{18Bertolami2008}.

To study the effect of the nonminimally coupled theory of gravity
in the cosmological context, we need to consider a flat, homogeneous
and isotropic universe by using the Friedmann-Robertson-Walker metric,
\begin{equation}
\label{modfreq6}
ds^{2}=-dt^{2}+a^{2}(t)(dr^2+r^2d\Omega^2),
\end{equation}
where $a(t)$ is the scale factor. We assume that the matter content
of the universe is described by a perfect fluid with the energy-momentum
tensor,
\begin{equation}
T_{\mu\nu}=(\rho+p)u_{\mu}u_{\nu}+pg_{\mu\nu},
\end{equation}
where $\rho$ is the energy density, $p$ is the pressure, and $u_{\mu}$
is the four-velocity satisfying $u^{\mu}u_{\mu}=-1$.

The $tt$ component of \eqref{modgreq1} yields the modified Friedmann equation \cite{13Bertolami2013}
\begin{equation}
\label{modfreq8}
H^{2}=\frac{1}{6F}[\rho+\lambda f_{2}\rho-6H\partial_{t}F+FR-f_{1}],
\end{equation}
where $H\equiv\dot{a}/a$ is the Hubble parameter and a dot stands
for a derivative with respect to $t$, and $F\equiv\kappa F_{1}+\lambda F_{2}\mathcal{L}$.

The nontrivial $\nu=t$ component of \eqref{modconseq1} yields the energy
conservation equation
\begin{equation}
\label{modfreq9}
\dot{\rho}+3H(1+\omega)=-\frac{\lambda F_{2}}{1+\lambda f_{2}}(\mathcal{L}+\rho)\dot{R},
\end{equation}
where $\omega=p/\rho$ is the equation of state of matter.

In the rest of our discussion, we will take $f_{1}(R)=R$ and $f_{2}(R)=(R/R_{0})^{n}$, where $R_0$ is the Ricci scalar today,
to study the dynamical behavior of the system. And we will express
the matter Lagrangian density in terms of the energy density $\mathcal{L}=-\alpha\rho$ \cite{13Bertolami2013}.
 We choose either $\alpha=1$ so that the matter
Lagrangian density $\mathcal{L}=-\rho$, or $\alpha=-\omega$ so that
the Lagrangian density  takes the form $\mathcal{L}=p$.

\section{Dynamical System}

Substituting $f_{1}(R)=R$, $f_{2}(R)=(R/R_{0})^{n}$ and $\mathcal{L}=-\alpha\rho$
into the modified Friedmann equation \eqref{modfreq8}, we obtain
\begin{equation}
\label{modfreq10}
H^{2}=\frac{1}{6F}[\rho+\lambda f_{2}\rho+6\lambda\alpha H\partial_{t}(F_{2}\rho)-\lambda\alpha F_{2}\rho R],
\end{equation}
and the energy conservation equation \eqref{modfreq9} can be rewritten in the form
\begin{equation}
\label{modfreq11}
\dot{\rho}+3H(1+\omega)\text{\ensuremath{\rho}}=\frac{\lambda F_{2}}{1+\lambda f_{2}}(\alpha-1)\rho\dot{R}.
\end{equation}
To study the cosmological dynamics of the model, we express the modified
Friedmann equation \eqref{modfreq10} into a dimensionless form by dividing it with
$H^{2}$, which results in
\begin{equation}
\label{modfreq12}
\begin{split}
1&=\frac{\rho}{6FH^{2}}+\frac{\lambda f_{2}\rho}{6FH^{2}}+\frac{\lambda\alpha\partial_{t}(F_{2}\rho)}{FH}-\frac{\lambda\alpha F_{2}\rho R}{6FH^{2}}\\
&=\frac{\rho}{6FH^{2}}+\frac{\lambda(1-\alpha n)f_{2}\rho}{6FH^{2}}+\frac{\lambda\alpha\partial_{t}(F_{2}\rho)}{FH}.
\end{split}
\end{equation}
In order to obtain the solutions of the field equations, we will employ
the appropriate dimensionless variables below,
\begin{gather}
\label{modfreq13}
x=\frac{\lambda(1-\alpha n)f_{2}\rho}{6FH^{2}},\\
\label{modfreq14}
y=\frac{\lambda\alpha\partial_{t}(F_{2}\rho)}{FH},\\
\label{modfreq15}
u=\frac{R}{6H^{2}}.
\end{gather}
We would like to indicate that the choices of these dimensionless
variables are particularly crucial for the dynamical system analysis
in the following, since they can accommodate the late time attractor solution in the gravity model as we will show below. The dimensionless Friedmann equation \eqref{modfreq12} then becomes $1 = \Omega + x + y$, where $\Omega \equiv \frac{\rho}{6FH^2}$ is the matter energy abundance, and we can regard $x+y$ as the effective energy abundance due to the coupling between matter and curvature. $u$ is the scalar curvature normalized to the Hubble parameter. Because of the constraint \eqref{modfreq12}, $\Omega$, $x$, $y$ and $u$ are not independent. To investigate the dynamics of the system, we need to find the equations of motion for $x$, $y$ and $u$.

Differentiating the modified Friedmann equation \eqref{modfreq10}, we obtain
\begin{gather}
\label{modfreq10a}
6\lambda\alpha\partial^{2}_{t}(F_{2}\rho)=3(1+\omega)(1+\lambda f_{2})\rho+6\lambda\alpha H\partial_{t}(F_{2}\rho)+12F\dot{H},
\end{gather}
where we have used \eqref{modfreq11} and $R=6(2H^{2}+\dot{H})$ given by the adopted metric \eqref{modfreq6}.

In terms of the variable quantities defined in \eqref{modfreq13}-\eqref{modfreq15}, we have
\begin{gather}
\label{modfreq10b}
\frac{\dot{R}}{RH}=\frac{3(1+\omega)\alpha nx+(1-\alpha n)yu}{\alpha nx}\cdot\frac{(1-\alpha n)(1-y-x)+x}{(n-1)[(1-\alpha n)(1-x-y)+x]-n(1-\alpha)x}\equiv \frac{s}{x},
\end{gather}

We are ready to derive the dynamical system using the variables above.
We will introduce a new ``time'' variable
$N=\ln a$, so that $dN=Hdt$. Using \eqref{modfreq10a} and \eqref{modfreq10b}, and differentiating
$x$, $y$ and $u$ with respect to $N$, respectively, we obtain
the following autonomous system,
\begin{gather}
\label{modfreq16}
\frac{dx}{dN}=\frac{1-\alpha n}{\alpha n}yu+xy-2xu+4x+s,\\
\label{modfreq17}
\frac{dy}{dN}=3(1+\omega)(1-x-y)+\frac{3(1+\omega)}{1-\alpha n}x+3y+y^{2}+2u-yu-4,\\
\label{modfreq18}
\frac{du}{dN}=\frac{us}{x}-2u^{2}+4u,
\end{gather}
where
\begin{equation}
\label{modfreq17a}
s=\frac{3(1+\omega)\alpha nx+(1-\alpha n)yu}{\alpha n}\cdot\frac{(1-\alpha n)(1-y-x)+x}{(n-1)[(1-\alpha n)(1-x-y)+x]-n(1-\alpha)x}.
\end{equation}
The complete dynamics of this cosmological model is described by these
three equations and the constraint \eqref{modfreq12}. The properties of the dynamical system \eqref{modfreq16}-\eqref{modfreq18} depend
on the values of the constants $\alpha$, $n$ and $\omega$. They
will in particular affect the existence and the stability of the fixed
points of the system.

In order to study the dynamics of the system, from $\frac{dx}{dN}=0$,
$\frac{dy}{dN}=0$ and $\frac{du}{dN}=0$, we can obtain the fixed
points of the system listed in Table \ref{tab.fix}. After finding all the possible
fixed points, we can further compute the eigenvalues and determine
their stabilities.

\begin{table}[ht]
\caption{The fixed points of the dynamical system \label{tab.fix}}
\begin{tabular}{|c|c|c|c|}
\hline
Fixed points & $x$ & $y$ & $u$\\
\hline
1 & $0$ & $1$ & $0$\\
\hline
2 & $0$ & $-1+3\omega$ & $0$\\
\hline
3 & $5-\frac{5}{n\alpha}$ & $-4$ & $0$\\
\hline
4 & $\frac{-2+5n\alpha+3\omega}{-1+n\alpha}$ & $\frac{1-4n\alpha-3\omega}{-1+n\alpha}$ & $0$\\
\hline
5 & $0$ & $0$ & $\frac{1}{2}(1-3\omega)$\\
\hline
6 & $1$ & $0$ & $\frac{-1+4n\alpha+3\omega}{-2+2n\alpha}$\\
\hline
7 & $1-\frac{1}{n\alpha}$ & $0$ & $2$\\
\hline
8 & $2-\frac{2}{n\alpha}$ & $-1$ & $2$\\
\hline
9 & $2-\frac{2}{n\alpha}$ & $3(1+\omega)$ & $2$\\
\hline
10 & $\frac{-2+4n\alpha+3\omega}{-1+2n\alpha}$ & $\frac{1-2n\alpha-3\omega}{-1+2n\alpha}$ & $\frac{n\alpha(-2+4n\alpha+3\omega)}{(-1+n\alpha)(-1+2n\alpha)}$\\
\hline
\end{tabular}
\end{table}

In order to determine the properties of the fixed points in the cosmological
context, we need to take into account some physical quantities in
cosmology.
An important parameter used in cosmology is the deceleration parameter,
which is defined as
\begin{equation}
q\equiv-\frac{\ddot{a}a}{\dot{a}^{2}}
\end{equation}
It can be expressed in the form $q=1-\frac{R}{6H^{2}}=1-u$. Since
our universe is expanding at an accelerated rate, we need to search
for a cosmological model with $q<0$, which puts the requirement that
$u>1$. The effective equation of state of the universe reads
\begin{equation}
\omega_{eff}\equiv-1-\frac{2\dot{H}}{3H^{2}}=\frac{1-2u}{3}
\end{equation}
and the accelerating expansion requires a fluid with negative
pressure, satisfying $\omega_{eff}<-1/3$.

Considering the transition from radiation domination to a later accelerated epoch, one barotropic fluid dominated saddle point which is for a radiation dominated era or a matter dominated era is sufficient to depict the dynamical process before the final attractor describing the universe acceleration \cite{19Boehmer2014}. The saddle point makes sure that
the Universe undergoes radiation or matter domination and is finally attracted to an accelerating expansion.
For the $\Lambda$CDM model \cite{19Boehmer2014}, the dynamical system is satisfied when there is only one saddle point in the deceleration and in
the late epoch the universe approaches the attractor solution undergoing an accelerating expansion, which is in consistent with
observations.

Now we turn to the analysis of the fixed points listed in Table \ref{tab.fix} in the
cosmological context. First we choose $\alpha=1$ in the matter
Lagrangian density, namely $\mathcal{L}=-\rho$. We show that our model is consistent with observations before
the late time acceleration. Thus in the radiation and matter dominated
eras, our model should be consistent with the standard observational cosmology, with $\omega=1/3$ and $q=1$ for the radiation era and
$\omega=0$ and $q=1/2$ for the matter era.  We obtain the stability of the fixed
points in the radiation era in Table \ref{tab.stab}. $u=0$ ensures the effective equation
of state $\omega_{eff}=1/3$, the deceleration parameter $q=1$ and
$\Omega=1$, indicating the radiation dominance. In the radiation
era, we would have a saddle point so that the universe will evolve
through this era. The fixed points R1 and R2 can meet this requirement. Their stability depends on the value of $n$.

For the matter dominant phase, the universe is filled
with a perfect fluid composed of nonrelativistic dust particles with
$\omega=0$.
The effective equation of state should be $\omega_{eff}=0$ and the
deceleration parameter needs to satisfy $q=1/2$ to ensure the matter
dominance with $\Omega=1$. The only possible fixed point candidate satisfying all these requirements in the
matter dominated era is M1 in Table \ref{tab.stab}. It is stable when $0<n<1$, or unstable
provided that $n<0$ or $n>1$. $n$ is the exponent of the curvature
scalar in $f_{2}(R)$. Therefore, the matter domination will not be the final state.


Now let us examine whether this model can provide us
a universe with late time accelerating expansion. Suppose the universe
is still filled with perfect fluid composed of nonrelativistic dust
particles with $\omega=0$. We find that the fixed point
D3 in Table \ref{tab.stab} can be stable when $n<0$ or $n>1$, which can serve
as an attractor solution. We get the effective equation of state
$\omega_{eff}=(1+3n)/(3-3n)$ and the deceleration parameter $q=(1+n)/(1-n)$.
Taking $n>1$ or $n<-1$, we can have the acceleration of the universe
with $q<0$. If $|n|$ tends to infinity, $q$ will approach -1 and the
effective equation of state will approach -1, so that the universe will
approach the asymptotic de Sitter phase. $|n|$ approaching infinity
is an extreme case, other values of $n$ can also accommodate the
late time acceleration in the model. Therefore, for $n>1$ or
$n<-1$, we find that the Universe undergoes radiation and matter domination which
are unstable fixed points in the phase space and is finally attracted to the accelerating expansion.

Then we choose $\alpha=-\omega$ so that the nature of the matter Lagrangian density has the form $\mathcal{L}=p$. We obtain the stability of the fixed points satisfying all requirements in Table \ref{tab.stabl}. In the radiation era, R2 is always a saddle point, while for R1, whether it is a stable or saddle point depends on the value of $n$. For the matter dominant phase, M1 is stable when $0<n<1$ and unstable when $n<0$ or $n>1$. For the late time era, the fixed points D3 is always a stable point, while D4 and D5 can be stable when $n<0$. For D1, D2 and D3, the matter density fraction $\Omega = 1 - x - y$ cannot be negative, so $n < 0$. We find that when $n<0$, neither radiation nor matter domination will be the final state, the universe will be attracted to the acceleration era. In the late time, the dominant component of the universe is nonrelativistic matter with $\omega \simeq 0$. For the fixed points D1, D2 and D3, one must set $|n| \gg 1$ in order to avoid the divergence in $x$. Similarly, for D4 and D5, though $\omega \simeq 0$ would not lead to divergence, the requirement $u > 1$ which guarantees the accelerated expansion of the universe also demands that $|n| \gg 1$. Therefore, when $\mathcal{L} = p$, we always need $|n| \gg 1$.

\begin{figure}
\begin{center}
\begin{tabular}{c}
\includegraphics[width=3.8in,height=2.9in]{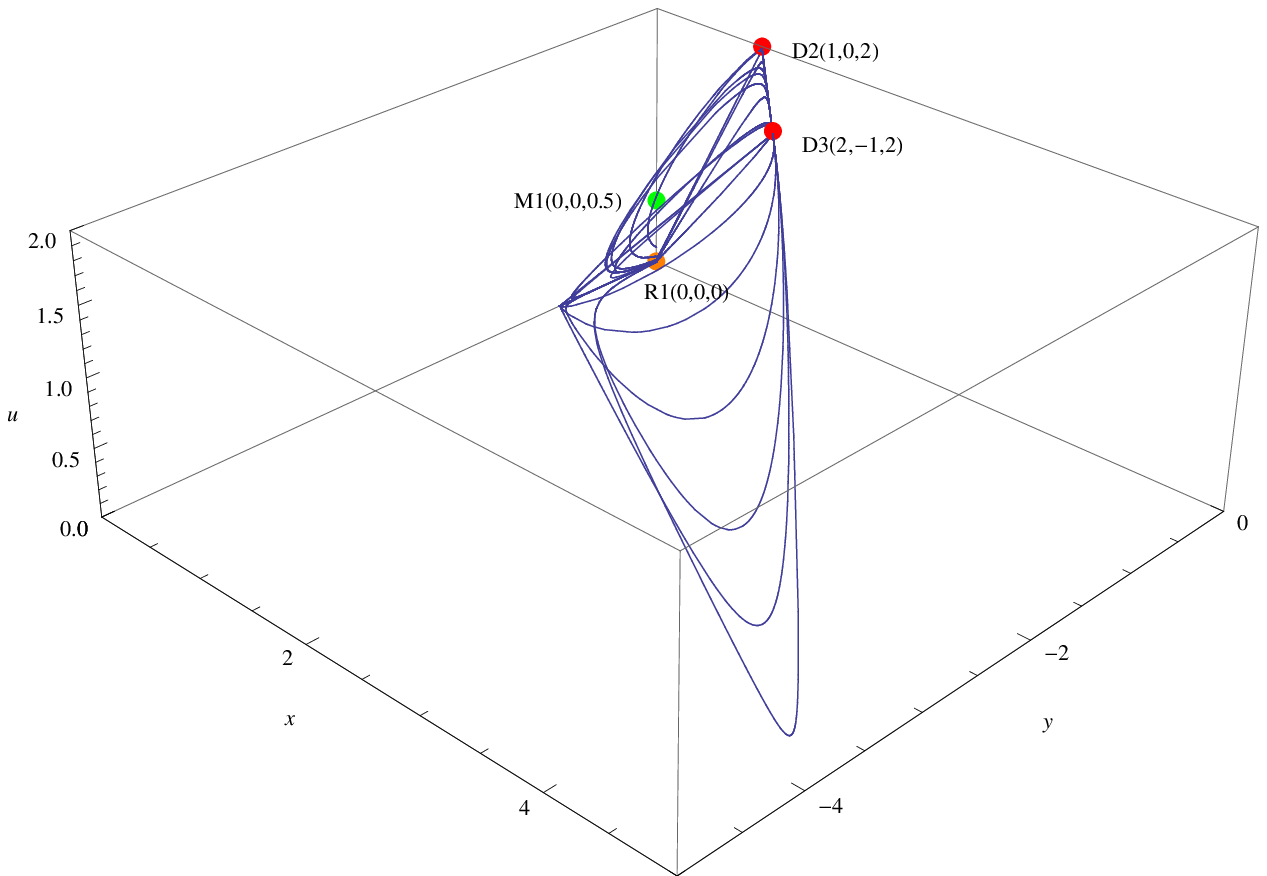}\nonumber
\includegraphics[width=2.6in,height=2.6in]{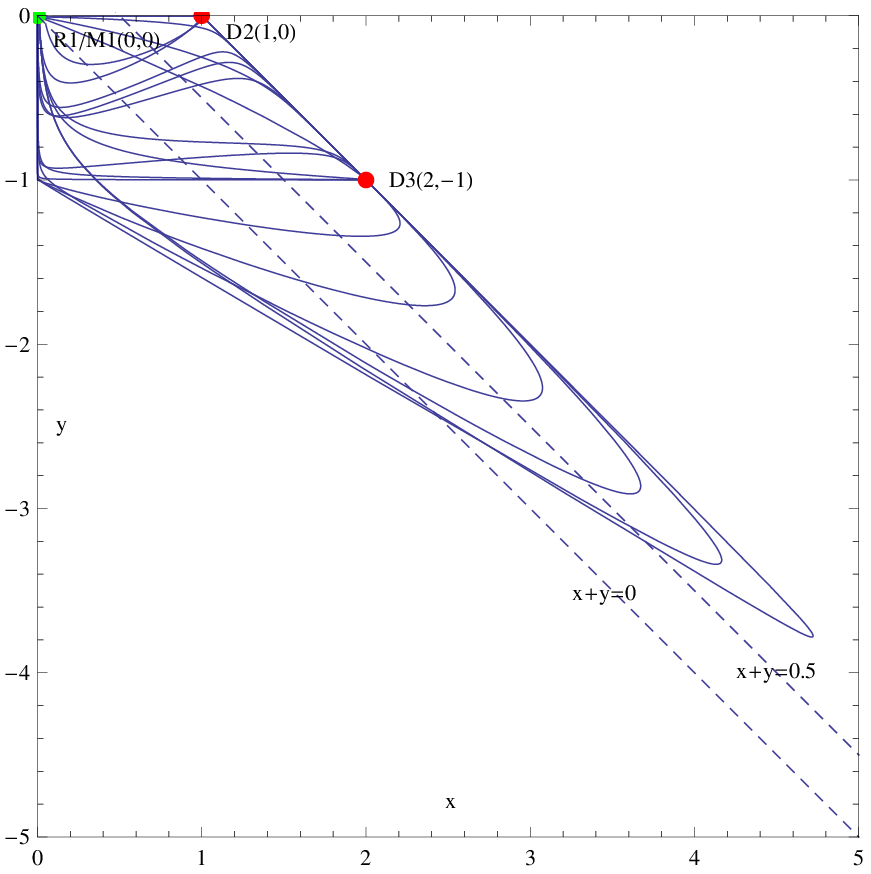}\nonumber\\
\end{tabular}
\end{center}
\caption{ Phase-space trajectories for the universe evolution described by the modified gravity with nonminimal gravitational
coupling to matter, where we choose $\mathcal{L}=-\rho$ and $n=-100$. We have also shown the projection plot. The point D3 is the  global attractor where all lines converge. }
\label{fig}
\end{figure}

In Fig. \ref{fig}, we show the phase-space trajectories for the universe evolution described by the modified gravity with nonminimally gravitational
coupling to matter. We see that the dynamics of the universe can evolve passing the saddle points such as R1 in the radiation era and D2 in the acceleration period and finally approaching the final attractor solution D3. For the unstable point in the matter dominated epoch M1, we see that trajectories for the universe evolution will detour around it but will not pass through it. This does not mean that the universe will not undergo a matter dominated epoch. In the two-dimensional projection figure we mark the matter dominated region between the dashed lines and see that all trajectories for the universe evolution pass through this matter domination region and finally converge to D3. Comparing with most of cosmological models, it is not so surprising that we here again find only one saddle point corresponding to the radiation dominated era in the deceleration of the universe. On the other hand the only unstable fixed point of the dynamic system in the matter dominated era deserve further studies. Whether this unstable point appears due to the choice of dynamical variables or it might indicate some problems of this model, for example whether the model can accommodate long enough matter dominated period for the structure formation, are questions we need to answer in the future.

We have shown that the modified nonminimally coupled gravity theory with $f_{2}(R)$ in terms
of the power law of the curvature scalar can accommodate the late time acceleration epoch of
the universe. Our result obtained through dynamical system analysis
confirms the solutions based on the studies assuming an specific ansatz
about the scale factor of the universe \cite{8Bertolami2010,9Bisabr2012}. The universe described by this model can finally enter the epoch of accelerated expansion.

\begin{table}[ht]
\protect\caption{The fixed points of dynamical system and their stability in different eras with $\mathcal{L}=-\rho$ \label{tab.stab}}
\centering{}%
\begin{tabular}{|c|c|c|c|}
\hline
\multicolumn{4}{|c|}{\textbf{Radiation era}}\\
\hline
\multicolumn{2}{|c|}{Fixed point {[}$x$, $y$, $u${]}} & $\omega=1/3$ & Stability\\
\hline
R1 & {[}$0$,$-1+3\omega$,$0${]} & {[}$0$,$0$,$0${]} & Stable: $0<n<1$,\\
 & & &
Saddle: $n<0$ or $n>1$\\
\hline
R2 & {[}$5-\frac{5}{n\alpha}$,$-4$,$0${]} & {[}$5-\frac{5}{n}$,$-4$,$0${]} & Stable: $-35+16\sqrt{5}\leq n<1$,\\
 & & &
Saddle: $n<0$ or $n>1$\\
\hline
\multicolumn{4}{|c|}{\textbf{Matter era }}\\
\hline
\multicolumn{2}{|c|}{Fixed point {[}$x$, $y$, $u${]}} & $\omega=0$ & Stability\\
\hline
M1 & {[}$0$,$0$,$\frac{1}{2}(1-3\omega)${]} & {[}$0$,$0$,$\frac{1}{2}${]} & Stable: $0<n<1$,\\
 & & & Unstable: $n<0$ or $n>1$\\
\hline
\multicolumn{4}{|c|}{\textbf{Late time acceleration era}}\\
\hline
\multicolumn{2}{|c|}{Fixed point {[}$x$, $y$, $u${]}} & $\omega \simeq 0$ & Stability\\
\hline
D1 & {[}$2-\frac{2}{n\alpha}$,$3(1+\omega)$,$2${]} & {[}$2-\frac{2}{n\alpha}$,$3$,$2${]} & Saddle for all\\
\hline
D2 & {[}$1$,$0$,$\frac{-1+4n\alpha+3\omega}{-2+2n\alpha}${]} & {[}$1$,$0$,$\frac{4n-1}{2n-2}${]} & Unstable: $\frac{1}{2}<n<1$,\\
 & & & Saddle: $n<\frac{1}{2}$ or $n>1$\\
\hline
D3 & $\left[\frac{-2+4n\alpha+3\omega}{-1+2n\alpha}, \frac{1-2n\alpha-3\omega}{-1+2n\alpha},
\frac{n\alpha(-2+4n\alpha+3\omega)}{(-1+n\alpha)(-1+2n\alpha)}\right]$ & $\left[2,-1,\frac{2n}{-1+n}\right]$ & Saddle: $0<n<1$,\\
 & & & Stable: $n<0$ or $n>1$\\
\hline
\end{tabular}
\end{table}

\begin{table}[ht]
\protect\caption{The fixed points of dynamical system and their stability in different eras with $\mathcal{L}=p $ \label{tab.stabl}}
\centering{}%
\begin{tabular}{|c|c|c|c|}
\hline
\multicolumn{4}{|c|}{\textbf{Radiation era}}\\
\hline
\multicolumn{2}{|c|}{Fixed point {[}$x$, $y$, $u${]}} & $\omega=1/3$ & Stability\\
\hline
R1 & {[}$0$,$-1+3\omega$,$0${]} & {[}$0$,$0$,$0${]} & Stable: $0<n<1$,\\
 & & &
Saddle: $n<0$ or $n>1$\\
\hline
R2 & {[}$5-\frac{5}{n\alpha}$,$-4$,$0${]} & {[}$5+\frac{15}{n}$,$-4$,$0${]} & Saddle for all\\
\hline
\multicolumn{4}{|c|}{\textbf{Matter era }}\\
\hline
\multicolumn{2}{|c|}{Fixed point {[}$x$, $y$, $u${]}} & $\omega=0$ & Stability\\
\hline
M1 & {[}$0$,$0$,$\frac{1}{2}(1-3\omega)${]} & {[}$0$,$0$,$\frac{1}{2}${]} & Stable: $0<n<1$,\\
 & & & Unstable: $n<0$ or $n>1$\\
\hline
\multicolumn{4}{|c|}{\textbf{Late time acceleration era}}\\
\hline
\multicolumn{2}{|c|}{Fixed point {[}$x$, $y$, $u${]}} & $\omega \simeq 0$ & Stability\\
\hline
D1 & {[}$2-\frac{2}{n\alpha}$,$3(1+\omega)$,$2${]} & {[}$2+\frac{2}{n\omega}$,$3$,$2${]} & Saddle for all\\
\hline
D2 & {[}$1-\frac{1}{n\alpha}$,$0$,$2${]} & {[}$1+\frac{1}{n\omega}$,$0$,$2${]} & Saddle for all\\
\hline
D3 & {[}$2-\frac{2}{n\alpha}$,$-1$,$2${]} & {[}$2+\frac{2}{n\omega}$,$-1$,$2${]} & Stable for all\\
\hline
D4 & {[}$1$,$0$,$\frac{-1+4n\alpha+3\omega}{-2+2n\alpha}${]} & {[}$1$,$0$,$\frac{4n\omega+1}{2n\omega+2}${]} & Stable: $n<0$,\\
 & & & Saddle: $n>0$\\
\hline
D5 & $\left[\frac{-2+4n\alpha+3\omega}{-1+2n\alpha}, \frac{1-2n\alpha-3\omega}{-1+2n\alpha},
\frac{n\alpha(-2+4n\alpha+3\omega)}{(-1+n\alpha)(-1+2n\alpha)}\right]$ & $\left[2,-1,\frac{2n\omega}{1+n\omega}\right]$ & Stable: $n<0$,\\
 & & & Saddle: $n>0$\\
\hline
\end{tabular}
\end{table}

\section{Radiation Temperature}

The thermodynamic interpretation of the generalized gravity theories
with geometry-matter coupling in the framework of the irreversible
thermodynamics of open systems was introduced in \cite{20Harko2014},
where the generalized conservation equations in the gravitational
theories were interpreted from a thermodynamic point of view as describing
the irreversible matter creation processes. According to the second
law of thermodynamics, the matter creation corresponds to an irreversible
energy flow from the gravitational field to the created matter constituents.
More discussions in this direction can be found in \cite{21Lobo2015,22Harko2015,23Harko2014}.

In this spirit, the energy conservation equation \eqref{modfreq9} reads
\begin{equation}
\dot{\rho}+3H(\rho+p)=\Gamma(\rho+p),
\end{equation}
where $\Gamma$ can be interpreted as the particle creation rate in
the thermodynamic explanations, which can be expressed in the form
\begin{equation}
\label{modfreq24}
\Gamma\equiv-\frac{1}{\rho+p}\frac{d}{dt}\ln[f_{\mathcal{L}}(R,\mathcal{L})](\mathcal{L}+\rho),
\end{equation}
where $f(R,\mathcal{L})=\kappa f_{1}(R)+[1+\lambda f_{2}(R)]\mathcal{L}$
in our model and the subscript in \eqref{modfreq24} stands for the derivative with
respect to $\mathcal{L}$.

Substituting $f_{\mathcal{L}}(R,\mathcal{L})=1+\lambda f_{2}(R)$
and $\mathcal{L}=-\alpha\rho$ into \eqref{modfreq24}, we have the particle
creation rate
\begin{equation}
\Gamma=\frac{\alpha-1}{1+\omega}\frac{d}{dt}\ln[1+\lambda f_{2}(R)].
\end{equation}
When the matter Lagrangian takes the form $\mathcal{L}=p$, during radiation era, the universe was filled with a perfect fluid composed of relativistic particles $\omega=\rho/p=1/3$ and $\alpha=-\omega=-1/3$, so that
\begin{equation}
\label{modfreq26}
\Gamma=-\frac{d}{dt}\ln[1+\lambda f_{2}(R)].
\end{equation}
For the matter Lagrangian  $\mathcal{L}=-\rho$, we have $\alpha=1$ which leads to the zero creation rate $\Gamma$.

With the help of the thermodynamic relations, we obtain the temperature
evolution of the particles as \cite{Lima2014}
\begin{equation}
\frac{\dot{T}}{T}=(\Gamma-3H)\frac{\partial p}{\partial\rho}.
\end{equation}
In the radiation dominated era, where radiation has pressure
$p=\rho/3$, one finds
\begin{equation}
\frac{\dot{T}}{T}=-\frac{\dot{a}}{a}+\frac{\Gamma}{3}.
\end{equation}
This equation can be rearranged as
\begin{equation}
T=T_{0}\left(\frac{a_{0}}{a}\right)\exp\left[-\frac{1}{3}\int_{t}^{t_{0}}\Gamma(t')dt'\right],
\end{equation}
or equivalently
\begin{equation}
\label{modfreq30}
T=T_{0}(1+z)\exp\left[\frac{1}{3}\int_{0}^{z}\Gamma(z')\frac{dt'}{dz'}dz'\right].
\end{equation}
where $a_{0}$ and $T_{0}$ are the present values of the scale factor
and of the radiation temperature, respectively.

Taking $\mathcal{L}=-\rho$, $\Gamma=0$, \eqref{modfreq30} returns to $T=T_{0}(1+z)$. While for the Lagrangian  $\mathcal{L}=p$, we have  nonzero creation rate \eqref{modfreq26}.
Substituting \eqref{modfreq26} into \eqref{modfreq30}, we have the temperature evolution
in the generalized gravity models with geometry-matter coupling
\begin{equation}
T=T_{0}(1+z)\left[\frac{1+\lambda f_{2}[R(z)]}{1+\lambda f_{2}[R(0)]}\right]^{-\frac{1}{3}}.
\end{equation}

Here we can clearly see the effect of the strength of coupling between matter
and geometry, which was hidden in the previous dynamics system analysis.
Taking $f_{2}(R)=(R/R_{0})^{n}$, we obtain the temperature of the
radiation
\begin{equation}
\label{modfreq32}
T=T_{0}(1+z)\left[\frac{1+\lambda[R(z)/R(0)]^{n}}{1+\lambda}\right]^{-\frac{1}{3}}.
\end{equation}
If there is no coupling between matter and geometry, $\lambda=0$,
so that $\Gamma=0$, which results in the temperature-redshift
relation $T=T_{0}(1+z)$. This is identical to the evolution of the radiation
temperature in the standard Einstein relativity in equilibrium and
is consistent with the observations. If there is coupling between
matter and geometry, the strength of the coupling will influence the
radiation temperature evolution. Besides the coupling, we see in \eqref{modfreq32}
that the power law parameter $n$ is also involved in the temperature
evolution if there is matter-geometry coupling.


\section{Observational constraint}

In this section we will constrain the parameters in our model. When $\mathcal{L}=p$, we have $\alpha=-\omega$. In the late time, the universe is filled with nonrelativistic matter particles with $\omega \simeq 0$, which requires $n$ to be negative and $|n| \gg 1$ as discussed above. This makes the numerical integration of the system difficult, thus we will only investigate the case $\mathcal{L}=-\rho$. To constrain the model parameters, we confront our model to observations. Theoretically we have shown that the power law parameter $n$ in the
modified gravity will influence the late time acceleration, since
it can modify the values of the effective equation of state and the deceleration
parameter. On the other hand, the analysis in the last section implies that the influence of the nonminimal coupling to the evolution of radiation temperature depends on the form of the matter Lagrangian. For $\mathcal{L} = -\rho$, the particle creation rate vanishes and the radiation temperature is not affected by the coupling. Thus we focus on the observations probing the expansion history of our universe.


We can clearly see in \eqref{modfreq10} that the luminosity distance is influenced by both the form of $f_2(R)$, which in our model is parameterized by the index $n$, and the coupling strength $\lambda$, due to the modification of Hubble parameter. From the dynamical system \eqref{modfreq16}-\eqref{modfreq18} we can calculate the H-z relation which deduce the theoretically luminosity distance relation. Therefore, in the analysis we take the measurements of luminosity distance from the type Ia supernova observations \cite{17Suzuki2012}. The $\chi_{L}^2$ of the luminosity distance reads
\begin{equation}
\label{modfreq34}
\chi_{L}^{2}(n,\lambda) = \overset{k}{\underset{i=1}{\sum}} \left[\frac{\mu_{obs}(z_{i})-\mu_{cal}(z_{i},n,\lambda)}{\sigma_{i}}\right]^{2},
\end{equation}
where $\mu_{obs}(z_{i})$ and $\mu_{cal}(z_{i},n,\lambda)$ are the
observed and theoretically calculated luminosity distance, respectively. $k$ represents the number of data points
and $\sigma_{i}$ is the uncertainty in the observed data.



\begin{figure}[ht]
\begin{centering}
\includegraphics[scale=0.76]{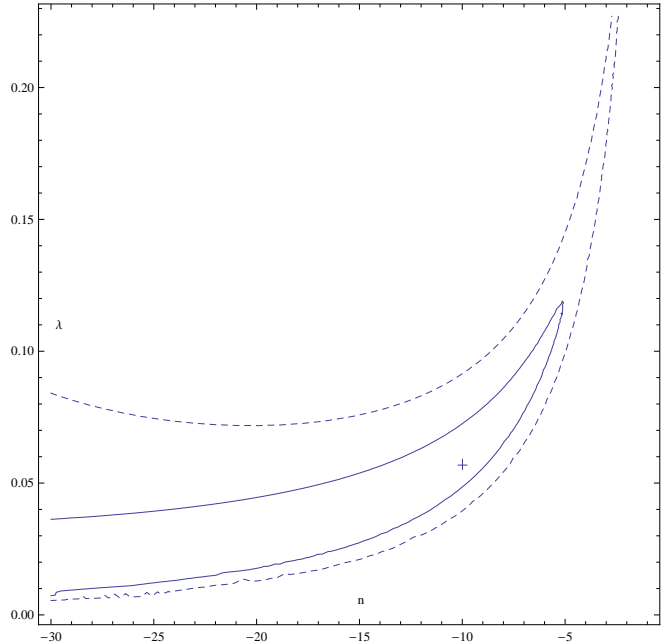}
\par\end{centering}

\protect\caption{Confidence contours. The solid and dashed lines represent the 68\%
and 95\% confidence regions, respectively. The `+' corresponds to $n=-10$ and $\lambda=0.057$. \label{fig.contour}}
\end{figure}

The likelihood function $L$ is defined as

\begin{equation}
L=\exp\left(-\frac{\chi_{L}^{2}}{2}\right),
\end{equation}
In the analysis we sample $n$ in the range $[-30, -1]$ and $\lambda$ in $[0.001,0.227]$. In Fig. \ref{fig.contour} we present the 68\% and 95\% confidence
contour plots of the normalized likelihood $L$ in the $(n,\lambda)$
plane.

\begin{figure}[ht]
\begin{centering}
\includegraphics[scale=0.77]{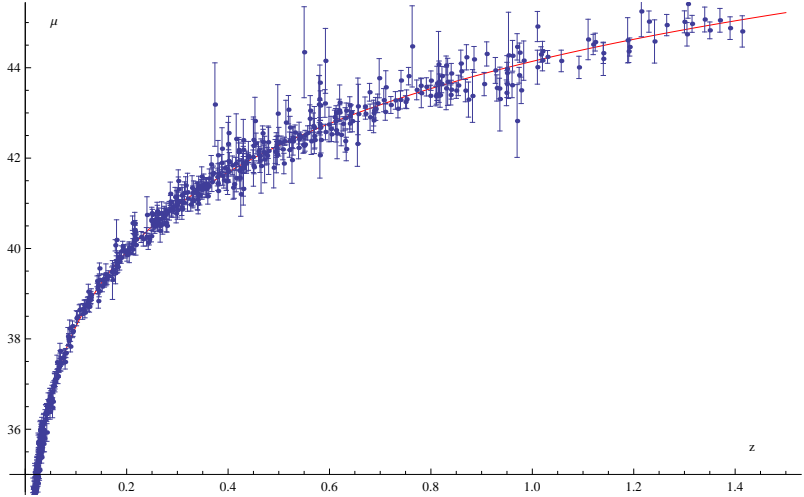}
\par\end{centering}

\protect\caption{The luminosity distance\textendash redshift relation. The red curve
represents the modified gravity model for $n=-10$ and $\lambda=0.057$. \label{fig.dist}}
\end{figure}

In Fig. \ref{fig.dist}, We plot the evolution of luminosity distance for $n=-10$ and $\lambda=0.057$, which lies in the $1\sigma$ confidence region in the parameter space, and compare them with the observational data sets. It is clear that the generalized gravity model with geometry and matter coupling can fit the observations well.



\section{Conclusions}

In this work, we have employed the dynamical system approach to investigate
the generalized gravity model with geometry-matter coupling. We
have learned that when matter is coupled explicitly to gravity the
Lagrangian densities describing perfect fluid composed of different
particles are not equivalent. Moreover in the cosmological context,
we have shown that different choices of Lagrangian density will apparently
result in different phases of the universe. This is one more evidence
showing that different Lagrangian densities are not equivalent when there is
coupling between matter and geometry.

In the nonminimally coupled theories of gravity, for the simple choices
of free functions $f_{1}(R)=R$ and $f_{2}(R)\sim R^{n}$,
we obtained a solution accommodating late time accelerating expansion of the universe by dynamical system approach. This is a systematic method to find solutions of the dynamics in general modified gravity theories without assuming particular forms of solution or initial conditions {\it a priori}. Our results confirmed previous solutions derived from some specific, {\it a priori} assumptions \cite{8Bertolami2010,9Bisabr2012}.

Following the thermodynamic understandings of the generalized
gravity theories with geometry-matter coupling, we further investigated the evolution of radiation
temperature in the model. Comparing the theoretical prediction of the model with the measurement of the luminosity distance, we constrained the model parameters $n$ and $\lambda$ when the Lagrangian density $\mathcal{L}=-\rho$. We found that the nonminimally coupled theory of gravity is compatible with current observations. Taking the model parameter $n<-5$,  the radiation dominated fixed point R1 is a saddle point, the matter dominated fixed point M1 is unstable and the accelerating fixed point D3 is stable,
so the model not only accommodates the usual radiation and matter dominant epoch, but also has an attractor corresponding to late time accelerated expansion. In \cite{27Bertolami2014}, the authors studied the nonminimally coupled generalized modified gravity model and obtained the conditions required for the absence of tachyon instabilities and ghost degrees of freedom. In particular, they found that one can avoid the Dolgov-Kawasaki instability if $n<0$, which is compatible with our constraint.

For the universe filled with the matter Lagrangian  $\mathcal{L}=p$, the dynamical analysis can also accommodate the universe to evolve from the radiation to matter dominated phase and finally settle down in the late time acceleration. However for this Lagrangian, when the late time universe is filled with nonrelativistic matter so that $\alpha=-\omega \rightarrow 0$, the exponent parameter $n$ is no longer free and it is required to approach negative infinity.

In \cite{26Amendola2007} it was argued that the $f(R)$ gravity with the inverse power law of the Ricci scalar is grossly inconsistent with cosmological observations although it can pass the supernova test. The reason for this $f(R)$ gravity was ruled out for viable cosmology is that it cannot supply the long enough matter dominated period for the structure formation. Whether this problem can be overcome in the model with nonminimally coupling between gravity and matter is an open question and worth pursuing.

\begin{acknowledgments}
We acknowledge financial supports from National Basic Research Program of China (973 Program 2013CB834900) and National Natural Science Foundation of China. Y. G. was also supported in part by the Natural Science Foundation of China under Grant No. 11475065, and the Program for New Century Excellent Talents in University under Grant No. NCET-12-0205.
\end{acknowledgments}

\end{document}